# Hyperelastic constitutive model for rubber-like materials based on the first Seth strain measures invariant

H. BECHIR[1], L. CHEVALIER[2, a], M. CHAOUCHE[3], K. BOUFALA[1]

[1] Laboratoire de Technologie des Matériaux et Génie des procédés (LTMGP), Université de Béjaia, Algérie
[2] Laboratoire de Mécanique (LaM), Université de Marne-la-Vallée, France
[3] Laboratoire de Mécanique et Technologie (LMT), ENS- Cachan, France

Corresponding author:

Professor Luc Chevalier

LaM- Marne la Vallée Universtity

5 bd. Descartes, Champs sur Marne

77454 Marne la Vallée Cedex 2 - France

luc.chevalier@univ-mlv.fr

# Hyperelastic constitutive model for rubber-like materials based on the first Seth strain measures invariant

H. BECHIR, L. CHEVALIER, M. CHAOUCHE, K. BOUFALA

## Abstract

The mechanical behaviour of isotropic and incompressible vulcanized natural rubbers (NR's) and that of quasi-incompressible carbon black filled vulcanized natural rubbers (NR 70) are considered both theoretically and experimentally. Based on the Seth strain measures in terms of the first invariant, an original form of the strain energy density function W is derived. This function is actually a generalisation of that of the neo-Hookean model and satisfies the hypothesis of the Valanis-Landel function. In the present study the analytical form of W is identified by using only simple tension test data (simple tension and simple planar compression). In our experiments, the two-dimensional field of in-plane homogeneous displacements is determined by using a home-developed image analysis cross-correlation technique. Our model is also identified using results taken from the literature in the case of (NR's) tested under simple tension, equibiaxial tension and pure shear. Comparison of numerical results with the experimental data indicates that the present model can characterise the hyperelastic behaviour of NR's and that of NR 70 for all the tested modes of deformation. Moreover, it seems to be valid over a wide range of deformation.

## 1. Introduction

Rubber-like materials are used in various engineering applications, like vibration-isolation devices, engine mounts, building and bridge bearings, vehicle door seals, tires, adhesive joints, etc.. Generally, these materials are characterised by high deformability and reversibility of deformation. Under purely static solicitations (without time effects), rubbers show hyperelastic behaviour. From a phenomenological point of view, the material is considered as a continuum and a strain energy density function W is postulated. W is a function of the local deformation gradient F. When the material is isotropic, W can be represented in terms of the three invariants $\{I_1, I_2, I_3\}$ that are defined through the principal stretches $\{\lambda_1, \lambda_2, \lambda_3\}$. It is common to assume that rubber-like materials are incompressible when they are not subjected to too large hydrostatic loadings. This assumption requires that $J = 1$ (where $J = \lambda_1\lambda_2\lambda_3$ ) and $I_3 = J^2$.

There are a number of reported strain energy density expressions in the literature for rubbers. The most widely used are those of Mooney-Rivlin (1940, 1948) and Ogden (1997). Descriptions of other models can be found in [Haines *et al.* (1979), Yamashita *et al.* (1993), O.H. Yeoh (1990, 1993), Lambert-Diani et al. (1999), Boyce *et al.* (2000), A.F.M.S. Amin *et al.* (2002)].

A pertinent model is the one that can lead to good agreement with experimental results for any stress state, with the same set of material parameters. Actually several material parameters are usually needed to take into account the non-linearity in the load-stretch relationships. Hopefully the number of material parameters should be related to the level of non-linearity, but not to the type of loading or loading state we would like modelling (for example, equibiaxial tension). The challenge for constitutive relationships is to use the results of one test (like simple tension test in homogeneous deformations) and this should require a simple material parameter identification method with limiting the number of parameters.

Existing models do not often allow describing the behaviour of rubbers with a wide range interval of strains, for example, simultaneously at small (under 10%) and large strains (over 100%) with the same set of material parameters. For incompressible materials such as rubbers, a $W(I_1, I_2, I_3=1)$ form is often less accurate than the Ogden form [Ogden (1997)] and leads to an unacceptable propagation of measurement error in the moderate strain region (e.g. 2-25%). For instance, Lambert-Diani *et al.* (1999) restricted their analysis in the region of large strains ($I_1 \geq 5$ and $I_2 \geq 5$). They developed a constitutive equation in terms of partial derivatives of the strain energy density function for rubbers and thermoplastic elastomers, and used the results data of two tests in homogeneous deformations (simple tension and equibiaxial tension) for the identification of the material parameters.

An important class of $W(\lambda_1, \lambda_2, \lambda_3)$ forms consists of those fulfilling $W = w(\lambda_1) + w(\lambda_2) + w(\lambda_3)$, where $w(\lambda_i)$ is the same function for each stretch component [Valanis *et al.* (1967), (1972)]. Ogden materials [Ogden (1997)] are of this separable form. Although accurate for rubber-like materials over a large range of stretch, this separable form is restrictive [Rivlin and Sawyers, (1976)].

Several molecular approaches were developed for modelling the mechanical behaviour of rubber-like materials, and corresponding continuum equivalent approaches were reported. For example, Arruda *et al.* (1993) proposed the eight-chain model that contains two molecular material parameters and used the simple tension test for material identification. To improve the equibiaxial tension results, these authors [(Boyce *et al.* (2000)] modified the Flory-Erman model (FE) by replacing the phantom strain energy by the eight strain energy. The "hybrid" strain energy density function contains another molecular parameter which is difficult to find out. Recently, Amin *et al.* (2002) modified the strain energy density function of Yamashita *et al.* (1993). The model was developed to describe the compression response of NR's. However, the behaviour of rubbers at large elastic strains may differ in tension and compression.

To improve constitutive descriptions of isotropic hyperelastic materials, we develop here a constitutive formulation based on the Seth strain measures invariant that is physically meaningful [Seth (1964)]. The non-linearity of the constitutive model is incorporated in the definition of the Seth strains invariant and the relationship between the strain energy density function and the strain invariant is linear. The advantage of the proposed model is that it is more reliable to describe the high level of non-linearity in the response of rubber-like materials. Moreover, we can use only one test in homogeneous strains (like simple tension test) for the identification of the material parameters. The simple linear method of least squares is used here for choosing the material parameters. The same set of material parameters are used for the prediction of other modes of deformation (pure shear, equibiaxial tension, simple compression and equibiaxial compression). The material parameters depend on the state of deformation. However, they are reliable to describe the response of rubbers in the wide range of strains, including moderate and large deformations.

In the next section, the constitutive equations are briefly reviewed using the theory of hyperelasticity. In section 3, the procedure of constructing a new strain energy based on Seth strains measures is reported. In section 4, the numerical prediction of our model is compared to experimental results taken from the literature [Treolar (1944) and Heuillet *et al.* (1997)] for (NR's) and to our experimental results concerning both NR’s and NR 70.

## 2. Review of the principle hyperelastic constitutive relationships

The strain energy density function W of a homogeneous material obeying the principle of objectivity is a function of the strain invariants $I_1$, $I_2$ and $I_3$ that are defined as the following:

$$I_1 = tr(C) = \sum_{n=1}^{n=3} \lambda_n{}^2 \qquad (1a);$$

$$I_2 = \frac{1}{2}\left[I_1{}^2 - tr(C)^2\right] = \sum_{n,m=1}^{n,m=3} (\lambda_n \lambda_m)^2 \qquad (1b);$$

$$I_3 = \det(C) = (\lambda_1 \lambda_2 \lambda_3)^2 \qquad (1c);$$

where $C=F^TF$ is the right Cauchy-Green tensor, tr(C) is the trace of C, F is the deformation gradient tensor, $\{\lambda_n\}$ are the principal stretch ratios and the symbol ($^T$) indicates the transpose of the corresponding tensor.

A stress-strain relationship can be derived from W. It may be given in terms of the first Piola-Kirchhoff stress tensor [Ogden, (1997)] $\pi=(\det F)\ \sigma\ (F^{-1})^T$, where $\sigma$ is the Cauchy stress tensor :

$$\pi = \frac{\partial W}{\partial F} = \frac{\partial W}{\partial I_1}\frac{\partial I_1}{\partial F} + \frac{\partial W}{\partial I_2}\frac{\partial I_2}{\partial F} + \frac{\partial W}{\partial I_3}\frac{\partial I_3}{\partial F} \qquad (2),$$

or equivalently :

$$\pi = 2\left[F \quad \frac{\partial W}{\partial I_1} + (I_1 F - FC)\frac{\partial W}{\partial I_2} + I_3 (F^{-1})^T \frac{\partial W}{\partial I_3}\right] \qquad (3).$$

For rubbers the condition of incompressibility is generally a good approximation. Under this condition, if one considers the problem of plane stress (simple tension, pure shear or biaxial tension) the constraint $I_3=1$ is identically true through the material. The strain energy density function is then a function of $I_1$ and $I_2$ only. The first Piola-Kirchhoff stress tensor reduces to a function of the partial derivatives ($\partial W/\partial I_1$) and ($\partial W/\partial I_2$) up to an arbitrary hydrostatic pressure p:

$$\pi = 2\left[ F\frac{\partial W}{\partial I_1} + (I_1 F - FC)\frac{\partial W}{\partial I_2} \right] - p(F^{-1})^T \quad (4)$$

The partial derivatives ($\partial W/\partial I_1$) and ($\partial W/\partial I_2$) correspond to the material parameters.

In some recently reported studies [O.H. Yeoh, (1990, 1993), Boyce *et al.* (2000), A.F.M.S. Amin *et al.* (2002)], it is shown that the strain energy density function can be written as a function of the first invariant $I_1(C)$ alone. Thus, these authors generalized the neo-Hookean model [Treolar, (1975)] and proposed the following decomposition of W:

$$W = W_{NH} + \gamma(I_1 - 3) \quad (5)$$

with $W_{NH} = C_1(I_1 - 3)$

$W_{NH}$ is the strain energy density function of the neo-Hookean model and $\gamma(I_1-3)$ a strain energy density function that depends on the state of deformation. Using higher order terms of $I_1$ in the term $\gamma(I_1-3)$ has been shown to well capture the deformation state at moderate to large deformations. Indeed, Yeoh (1990) proposed a polynomial form of $\gamma(I_1-3)$, which is actually a truncated form of the well-known Mooney-Rivlin energy density function (1940, 1948):

$$W_Y = C_{10}(I_1 - 3) + C_{20}(I_1 - 3)^2 + C_{30}(I_1 - 3)^3 \quad (6)$$

This function was later modified by adding an exponential term [Yeoh (1993)] to improve low strain accuracy:

$$W_Y = C_{10}(I_1 - 3) + C_{20}(I_1 - 3)^2 + C_{30}(I_1 - 3)^3 + \frac{\alpha}{\beta}[1 - \exp(-\beta(I_1 - 3)] \quad (7)$$

where $C_{10}$, $C_{20}$, $C_{30}$, $\alpha$, and $\beta$ are material parameters.

An alternative high order $I_1$ model has been proposed recently by Boyce *et al.* (2000) :

$$W_{AB} = C_{10}\left(I_1 - 3\right) + \sum_{n=2}^{\infty} C_n\left(I_1^{\,n} - 3^n\right) \quad (8)$$

This model is equivalent the continuum version of the 8-chain molecular model [Boyce *et al.* (1993)]. To improve predictions of the model in equibiaxial tension at small strain region, Boyce *et al.* (2000) modified the Erman-Flory model ($W_{FE}$) by replacing the strain energy of the phantom Gaussian chains ($W_{ph}$) by the 8-chain model (Eq. (8)). In the Flory-Erman approach the elastic strain energy of the network is considered to be the sum of phantom and constraint contributions:

$$W_{FE} = W_{ph} + W_c \quad (9)$$

where $W_c$ is the strain energy density function of the constraints. We have :

$$W_c = C_{10}\sum_i \left[B_i + D_i - \ln\left(1 + B_i\right) - \ln\left(1 + D_i\right)\right] \quad (10)$$

with $B_i = k^2\left(\lambda_i^{\,2} - 1\right)\left(\lambda_i^{\,2} + k\right)^{-2}$ and $D_i = \lambda_i^{\,2}k^{-1} - B_i$. k is a molecular parameter that depends on the state of the network. The "hybrid" model proposed by Boyce *et al.* (2000) is expressed as:

$$W_{AB}^{H} = W_{AB} + W_c \quad (11)$$

It is to be noted that Boyce *et al.* (2000) did not give the equivalent continuum model to the constraints strain energy density function $W_c$ in Equation (11).

In order to describe the softening effects at low stretches, Amin *et al.* (2002) suggested to add two coefficients ($C_4$ and M=0.25) for the strain energy density function to that initially proposed by Yamashita *et al.* (1993).

$$W_{AMIN} = C_{10}\left(I_1 - 3\right) + \frac{C_3}{N+1}\left(I_1 - 3\right)^{N+1} + \frac{C_4}{M+1}\left(I_1 - 3\right)^{M+1} \quad (12)$$

where $C_{10}$, $C_3$, $C_4$, M, and N are material parameters with $N \geq 1$ and $0 \leq M \leq 1$.

This new formulation was expected to describe the data over entire range of deformation in terms of stress-strain relationship and for all modes of deformation for rubbers (NR's and HDR's). Amin *et al.* (2002) used the results of simple compression test data for the identification of the material parameters. The model was tested experimentally by comparison with simple compression stress-strain measurements performed up to large strains. However significant experimental/theoretical deviations still remained.

We conclude that the strain energy density function can be expressed as the sum of the polynomial function of $(I_1-3)^n$, with the values of n are integers, and a non linear function of $(I_1-3)$. Using classical invariants ($I_1(C)$ and $I_2(C)$) leads to the models, which require a specific and a coast experimental apparatus, to realize homogeneous deformations tests (i.e. biaxial test), in order to identify the material parameters.

In the present study, we propose a generalization of the strain energy density function (Eq. (5)). It has a polynomial form, and consists of a sum of the neo-Hookean model and a non- linear function of new invariants ($I_{(n)}(C)$) obtained with considering n-measure of strain. The idea of a generalized strain measure is not new. Indeed, Blatz *et al.* (1974) applied successfully the n-measure of strain for several rubber-like materials, considering "n" as a material parameter and must be determined by experiment. It is generally not integer. However, strain is indeed defined as purely kinematics variable, and it is not a material property. This problem has recently been pointed out by K. Farahani *et al.* (2004). They used the concept of conjugate stress-strain for the generalization of Hooke's law in finite elastic deformations, where the values of n are integers (i.e. n=-2, -1, 0, 1, 2). From the phenomenological point of view, one can use the strain measure as material property [Chang et *al.* (1976)] for applications in rubber engineering. In this paper, the values of n are integers (n=1, 2, 3). The advantage of our approach is to introduce the high level of non linearity of the constitutive equation in the definition of the invariants $I(n)(C)$ . Hence, the strain energy density function is expressed as a polynomial form of $(I(n)-3)^m$ . The material parameters will be inferred by a simple identification method (square linear method), using only the results of the simple tension test data. Introducing Eq. (5) into Eq. (4), we obtain the following constitutive relationship:

$$\Pi=2F\frac{dW}{dI_1}-pF^{-T} \tag{13}$$

# 3. Strain energy density function based on Seth strain measures

## *3.1 Seth strain measures*

The stress-strain relationships for rubbers are non-linear in the entire range of extensibility. Instead of including the non-linearity aspect in the constitutive relation between the strain energy density function, it is possible to introduce the non-linearity in the strain invariant. A way to do that is through the Seth-Hill strain measures $E^{(n)}$, defined as:

$$E^{(n)}=\frac{1}{2n}\left(C^n-I\right);\ \text{if } n\neq 0 \tag{14a}$$

$$E^{(n)}=\frac{1}{2}\,Ln\left(C\right);\ \text{if } n=0 \tag{14b}$$

The polar decomposition theorem states that F may be uniquely decomposed:

$$F=RU=VR \tag{15}$$

U and V are the right and left stretch tensors respectively. They are positive, definite and symmetric tensors. R is the rotation matrix. The eigenvalues of U and V are the principal stretches $\{\lambda_1, \lambda_2, \lambda_3\}$. The spectral theorem decomposition leads to :

$$U=\sum_{i=1}^{i=3}\lambda_i N_i \otimes N_i \tag{16}$$

where $N_i$ are the orthonormal eigenvectors of U. Combining Eqs. (7a) and (9), the Seth-Hill strain measures tensors $E^{(n)}$ can be expressed as a function of the principal stretches $\{\lambda_i\}_{i=1,2,3}$:

$$E^{(n)}=\frac{1}{2n}\left(U^{2n}-I\right) \tag{17a}$$

$$E^{(n)}=\frac{1}{2n}\left[\left(\sum_{i=1}^{i=3}\lambda_i^{2n} N_i \otimes N_i\right)-I\right] \tag{17b}$$

We can then choose the first invariant of Seth–Hill strain measures to generalize the strain energy density function for rubbers (Eq. (5)):

$$I_{(n)}\left(E^{(n)}\right)=tr\left(E^{(n)}\right)=\frac{1}{2n}\left(tr\left(C^{n}\right)-3\right)=\frac{1}{2n}\left(\lambda_1^{2n}+\lambda_2^{2n}+\lambda_3^{2n}-3\right) \tag{18}$$

Inserting Eq. (10) into Eq. (5) and expanding W in powers of $I_{(n)}$ [$E^{(n)}$ ], we obtain:

$$W=\sum_{n=1}^{\infty}\sum_{r=1}^{\infty}\alpha_n^r I_{(n)}\left(E^{(n)}\right)^r=\sum_{n=1}^{\infty}\sum_{r=1}^{\infty}C_n^r\left(\lambda_1^{2n}+\lambda_2^{2n}+\lambda_3^{2n}-3\right)^r \tag{19a}$$

For some material parameters $C_n^r$, we see that our model (Eq. (19a)) can be expressed in a form similar to that in Eq. (5):

$$W=W_{NH}+\sum_{n=2}^{\infty}\sum_{r=2}^{\infty}C_n^r\left(\lambda_1^{2n}+\lambda_2^{2n}+\lambda_3^{2n}-3\right)^r \tag{19b}$$

The first advantage of the present model (Eq. (19)) is that it satisfies the special form of Valanis-Landel (VL) function [Valanis *et al.* (1967, 1972)]. The VL assumption states that a strain energy density function for rubbers can be written as a sum of independent functions of the principal stretches:

$$W=w(\lambda_1)+w(\lambda_2)+w(\lambda_3) \tag{20}$$

Bradley *et al.* (2001) showed that Ogden model (1972), which is a special form of VL function, gives reasonable means for estimating the three-dimensional strain energy density when only simple tension data are used. Indeed, under the assumption of VL function, for an incompressible material, simple compression (SC) is equivalent to simple tension (ET) and equibiaxial tension (ST) is equivalent to equibiaxial compression (EC).

## 3.2 Stress-strain relationship for incompressible rubbers

We now examine the constitutive equations inferred from our strain energy density function for different deformation fields. Substitution of Eq. (19) into Eq. (6) gives :

$$\pi = 2F\sum_{k=1}^{k=\infty}\frac{\partial W}{\partial I_{(k)}}\frac{\partial I_{(k)}}{\partial I_1} - p\left(F^{-1}\right)^T \tag{21}$$

Equation (21) can now be worked out for different types of deformation fields, in particular those involved in our experiments.

### *3.2.1 Simple tension (ST) or simple plane compression (SC)*

In simple tension and simple compression, we have respectively:

$\lambda_2=\lambda$, $\lambda_1=\lambda_3=\lambda^{-1/2}$, $\pi_2=\pi^{ST}$, $\pi_1=\pi_3=0$, and $\lambda_1=\lambda_2=\lambda$, $\lambda_3=\lambda^{-2}$, $\pi_1=\pi_2=0$, and $\pi_3=\pi^{SC}$.

Hence, from Eq. (20),

$$\pi^{ST} = w^{/}(\lambda) - \frac{1}{\lambda^{\frac{3}{2}}} w^{/}\left(\frac{1}{\sqrt{\lambda}}\right) \tag{22}$$

and,

$$\pi^{EC} = w^{/}\left(\frac{1}{\lambda^2}\right) - \lambda^3 w^{/}(\lambda) \tag{23}.$$

For an incompressible material, the state of deformation in simple tension (ST) and simple compression (SC) is the same (see, fig. (1)).

Through a sample transformation, one can show that the simple plane compression stress-strain can be written as a function of the simple tension:

$$\pi^{SC}(\lambda) = -\pi^{ST}(x) \text{ , with } \lambda = \frac{1}{x^2} \tag{24}$$

*3.2.2 Equibiaxial tension (ET) or equibiaxial compression (EC)*

In equibiaxial tension, we have : $\lambda_1=\lambda_2=\lambda$, $\lambda_3=\lambda^{-2}$, $\pi_1=\pi_2=\pi^{ET}$ and $\pi_3=0$. Using Eq. (20), we obtain:

$$\pi^{ET}=w'(\lambda)-\frac{1}{\lambda^3}w'\left(\frac{1}{\lambda^2}\right) \qquad (25)$$

The state of deformation in the equibiaxial tension and the equibiaxial compression tests are equivalent (see, fig. (21)).

Actually, for an incompressible material, we can generate the stresses for equibiaxial compression from stresses and stretch of equibiaxial tension data:

$$\pi^{EC}(x)=-2.\lambda^{-\frac{3}{2}}\pi^{ET}(\lambda), \text{ with } x=\sqrt{\lambda} \qquad (26)$$

*3.2.3 Pure shear (PS)*

In pure shear, we have : $\lambda_1=\lambda$, $\lambda_2=1$, $\lambda_3=\lambda^{-1}$, $\pi_1=\pi^{PS}$ and $\pi_2=\pi_3=0$,

Eq. (20) leads to:

$$\pi^{PS}=w'(\lambda)-\frac{1}{\lambda^2}w'\left(\frac{1}{\lambda}\right) \qquad (27)$$

# 4. Experimental

## *4.1. Testing set-up*

To test the validity of our strain energy density function (Eq. (19)), we undertake experiments involving a state of homogeneous strain on NR 70. The experiments consist of simple tension and simple plane compression. We also test our model using experimental results data taken from the literature [Treolar, (1944), and Heuillet *et al.* (1997)] for incompressible NR's.

Experimental tests are performed in simple tension and simple planar compression. These tests are carried out on a Deltalab testing machine. The specimens are rectangular (80x40x2 mm$^3$) in the case of simple tension tests and cylindrical (29 mm diameter and 13 mm height) for the compression tests.

Displacements are applied along the vertical axis of the specimen and a load cell measured the corresponding normal force. The in-plane deformation field is followed using a CCD camera. A home-developed digital data processing technique [Chevalier *et al.* (2001)] is used to analyse the displacement field during the loading of the specimen. This technique enables validation of the strain field homogeneity during the test. This homogeneity is easy to obtain during a tension test : on Fig.3a we can observe deformed grid during a tension test on NR and Fig.3b shows longitudinal and transversal displacement components. Both results clearly represent an uniform strain field.

The compression set up is outlined in Fig. 4a. In order to reduce the friction between the sample and the platens and thereby to ensure a homogeneous deformation, a low viscosity lubricant is inserted between the platens and the specimen. Contour line of the two components of the transversal displacement during compression are plotted on Fig.4b, once again the homogeneity is establish: the lubricant enable the specimen to slide on the lower platen and we can assume that sliding also occurs on the upper one.

### 4.2 Mechanical behaviour of NR 70

The tension and compression tests are carried out at room temperature (T ≅ 20ºC) and low strain rate (i.e. $\dot{\varepsilon} = 0.0045 s^{-1}$). The specimens are conditioned by six loads to remove the influence of the Mullins effect [Mullins *et al.*(1965)], and to insure repeatability in the tests. Figure 5a represents the mechanical behaviour of a typical NR 70 specimen in simple tension test. The elongation is inferred from the local strain which is measured using Correlli$^{GD}$. The Cauchy tension stress $\sigma$ (determined from the tension force F reduced to the section S) is plotted versus strain $\varepsilon$ ($\varepsilon = \ln\lambda$). We can observe non-linearity of the stress-strain curve at small strains ($\varepsilon \geq 0.097$). This can be attributed to non-Gaussian effects due to the limited chain extensibility.

There is a significant difference between our results and those obtained on CB filled NR by James *et al.* (1975). Nevertheless their material is actually different

The evolution of the tangent Young's modulus (i.e. $\overline{E}=\lambda\frac{d\sigma}{d\lambda}=\frac{d\sigma}{d\varepsilon}$ ) as a function strain, $\varepsilon$, is represented in Fig. 5b. For large deformations ($\varepsilon \geq 0.58$), we can observe a huge increase of the tangent Young's modulus. This may be due to two possible phenomena: (i) strain-induced crystallization, where the crystallites may act as additional reinforcement, and/or (ii) deformation of carbon black (CB) particles fillers.

To check the incompressibility assumption of the material, we use the results obtained in the case of simple tension. Hence, we introduce the Poisson ratio $\nu$ defined as:

$$\lambda_2 = \lambda_1^{-\nu} \tag{28}$$

with $\lambda_2=1+\varepsilon_{22}$ and $\lambda_1=1+\varepsilon_{11}$. $\varepsilon_{11}$ and $\varepsilon_{22}$ are respectively the transversal and longitudinal strain. These quantities are determined using Correlli$^{GD}$ .

In Fig. 6, we represent the evolution of $\lambda_2$ as a function of $\lambda_1$ for a simple tension test. The experimental value of the Poisson ratio is $\nu=0.48$. This value is close to 0.5 indicating that the material is quasi-incompressible.

The main difficulty is to insure homogeneous strain field in the specimen during the compression test. To overcome this difficulty, a low viscosity lubricant is used between the platen and the specimen. This leads to the radial expansion of the strain in the plane (1, 2) of the specimen when the material is constrained along the 3-axis. The assumption of incompressibility is checked in plane compression tests. The elongation is obtained from the transversal elongation measured by image analysis as function of the elongation calculated from the displacements measurements of the traverse testing machine. We have: $\lambda_2=\lambda_1=\lambda=1+\varepsilon$ ($\varepsilon=\varepsilon_{11}=\varepsilon_{22}$), and $\lambda_3=1-(\Delta h/h_0)$ with $\lambda=(\lambda_3)^{-\nu}$. We found that the Poisson ratio $\nu$ is equal to 0.26, which is quite different from the one determined in the simple tension test ($\nu$=0.48). The rubber-like materials are assumed to be isotropic. That is, the value of the Poisson ratio is the same throughout the medium. Fig. 7 illustrates the plot of the transversal elongation determined using a value of the Poisson ratio equal to 0.48 for plane compression test, and, is compared to that obtained with image analysis measurements. This difference can be explained by the presence of friction between the sample surface and the platen. Consequently, we use the experimental results data of simple tension test for modelling the response of NR 70, and simulated the response of the material in all modes of deformation.

In Fig. 8, the Cauchy stress in simple plane compression is plotted versus the deformation $\varepsilon$ or elongation for a range of elongation varying from 0.34 to 1. During the test, the polymer chains in the material stretch freely in all directions within a plane perpendicular to the load axis. A uniform radial expansion bulging of the specimen, it is evidence, that the strain field is homogeneous. However, the effect of the friction between the specimen and the platen introduce a correction on the measures (force and displacements), these values may be overestimated. The stress-strain plot in tension and compression reveals a continuity at $\varepsilon=0$ (cf. Fig. 8b) and linear behaviour at small strains. For ($\varepsilon\leq -0.176$), a shape upturn is observed, which may be explained by the non-Gaussian nature of the network.

### 4.3 Identification procedure and validation

Now, we present the identification method of the strain energy density function W (Eq. (19)). The material parameters are determined here using a least squares regression analysis of the experimental data in simple tension. This is easily implemented, using commercial package MATLAB. The material parameters $C_n$ are concomitant and dependent on the state of deformation.

Firstly, we assume that the strain energy density function W is neo-Hookean ($W=W_{NH}=C_1^1(I_1-3)$), the parameter $C_1^1$ is determined using simple tension data. In the second step, we evaluate the number of terms ($C_n$, n is integer, n=1, 2, 3 and r=1, 2,3) needed to approximate the strain-energy density function (Eq. (19)). The constants are listed in table II. The constants chosen for simple tension are used in the prediction of the material behaviour in the other deformation fields (pure shear and equibiaxial tension).

Figure 9 shows the experimental results of simple tension and simple plane compression tests on NR 70. The prediction of our model is excellent for simple tension test and, the stresses simulated for pure shear and equibiaxial tension tests are acceptable results. However, our model fails to predict the mechanical behaviour of the material in simple plane compression test. The reason of this discrepancy may be due to the following: the stresses calculated from simple tension data results test are not equal to the stresses measured in simple plane compression test, then the simple plane compression test is not perfect.

The results of Treolar (1944) and Heuillet *et al.* (1997) are shown in Fig. 10 and Fig. 11 respectively. The results for equibiaxial compression and simple compression are calculated respectively, from the experimental results data of equibiaxial tension and simple tension tests. It is shown in fig. 9 that four parameters ($C_1^1, C_1^2, C_2^1, C_2^2$) for the strain energy density given by Eq. (19) lead to quite good agreement with the experimental results of Treolar (1944). Significant discrepancy is observed only for large stretches $\lambda>4$ in the case of equibiaxial tension. This can be attributed to the chosen number of material parameters.

Ogden (1997) obtained an excellent fit with a six parameter model. We need to determine an optimum between the number of materials parameters and the complexity of the strain energy density function in terms of the invariants $\left(I_{(n)}-3\right)^m$.

Figure 11 shows that our model ( $C_1^1, C_2^1, C_3^1, C_1^2, C_2^2, C_3^2$ ) fit well the simple tension and pure shear results data of Heuillet *et al.* (1997). Our simulation of an equibiaxial tension test leads to good agreement with experimental results in the domain of moderate strain, but we obtain some discrepancy at large stretches $\lambda>3$. This may not be due to our approach, but rather would be induced the limitation of to the positive values of n. That is, Eq. (19) can be extended to minus values of n (i.e. n=-3, -2, -1, 1, 2, 3), with $0< r \leq n$.

## 5. Conclusion

In the present study a new strain energy-density function for isotropic and incompressible rubber-like materials is developed based on the Seth-Hill strains measures, which turned out to be a generalization of the neo-Hookean model. With regard to the $W(\lambda_1, \lambda_2, \lambda_3)$ approach of Ogden (1997), we consider the predictive capability of their approach to be equal to ours for materials that satisfy the Valanis-Landel (VL) hypothesis. Indeed, the advantage of our model is the relatively few material parameters to be determined simply. For instance, it was shown that four parameters were enough to lead to good comparison with experimental results of Treolar (1944), and with six parameters model, we could fit pretty well the experimental results of Heuillet *et al.* (1997). Many forms of $W(I_1, I_2)$ have been proposed for rubbers with some good predictions of experimental data. However, for wild stretch ($I_1< 5$ or $I_2< 5$), equibiaxial is singular for all $I_1$ and $I_2$ , and the decoupling effect of $I_1$ and $I_2$ in $W(I_1, I_2)$ [cf.

Diani *et al*. (1999)] is not possible, which explained the limitations of the W($I_1$, $I_2$) approach for moderate strains. For incompressible rubbers that satisfy the VL function, it is possible to obtain results in terms of the stress-strain data without making experimental tests. Indeed, we can generate all modes of deformation by using the simple tension and equibiaxial tension results tests data. We showed that our model can characterize pretty well the mechanical behaviour of rubbers in various deformation fields using the same set of material parameters determined from simple tension test.

## List of figures and tables

**Figure 6**: Logarithm of the simple lateral ratio versus Logarithm of the longitudinal extension.

**Figure 7**: Comparison of the lateral elongation measured with the image analysis in simple plane compression and, the same results calculated from simple tension test, the values of the slope is different from the identified Poisson ratio.

**Figure 8(a)**: Stress-strain of NR 70 in and simple plane compression, elongation calculated from the displacements of the traverse testing machine.

**Figure 8(b)**: Stress-strain response of NR 70 in simple plane compression and simple tension tests.

**Figure 9**: Stress-strain results measured in simple tension and in simple plane compression on NR 70, and the predictions of our Model.

**Figure 10**: Comparison of the predicted results of our Model to the experimental results of Treolar (1944) in the case of 8% sulphur rubber.

**Figure 11**: Comparison of the prediction of our Model to the experimental results of Heuillet *et al.* (1997) in the case of NR (Natural Rubber).

**Table I** : Composition of Carbon Black (C.B.)/Natural Rubber (NR) composites, the amount of C.B. 70% .

**Table II**: Material parameters for rubbers (NR's) and NR 70.

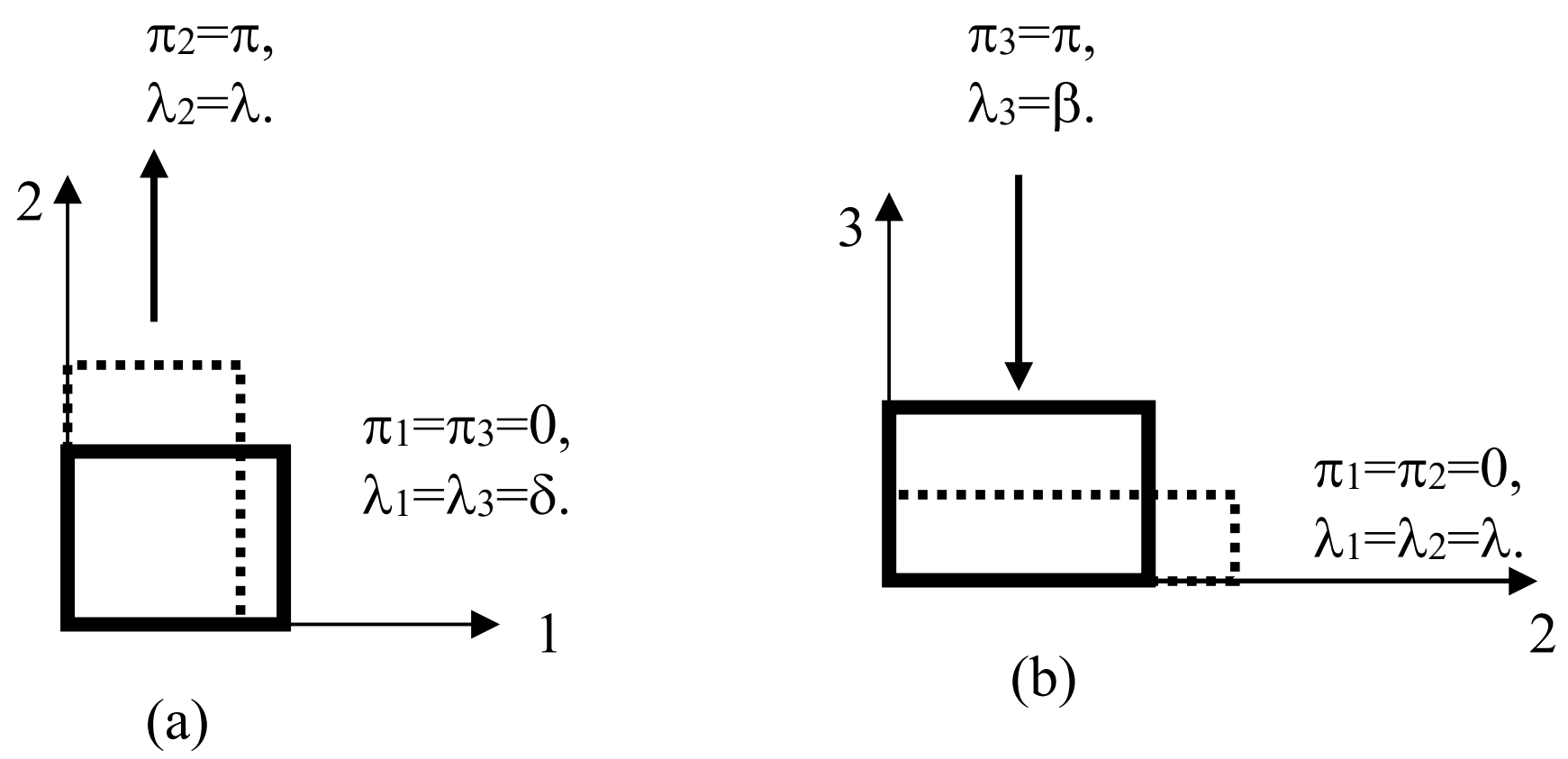

**Figure 1**: The state of deformation in our tests
(a): Simple tension
(b): Simple plane compression

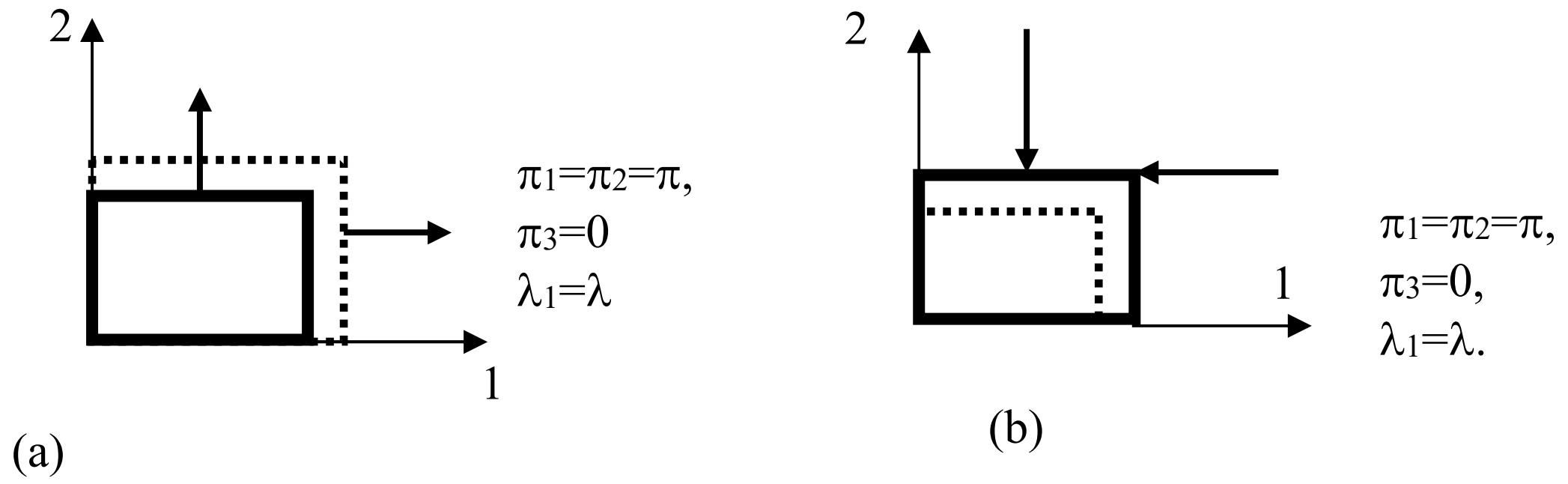

**Figure 2**: State of deformation in equibiaxial tension and equibiaxial compression
(a): Equibiaxial tension,
(b): Equibiaxial compression.

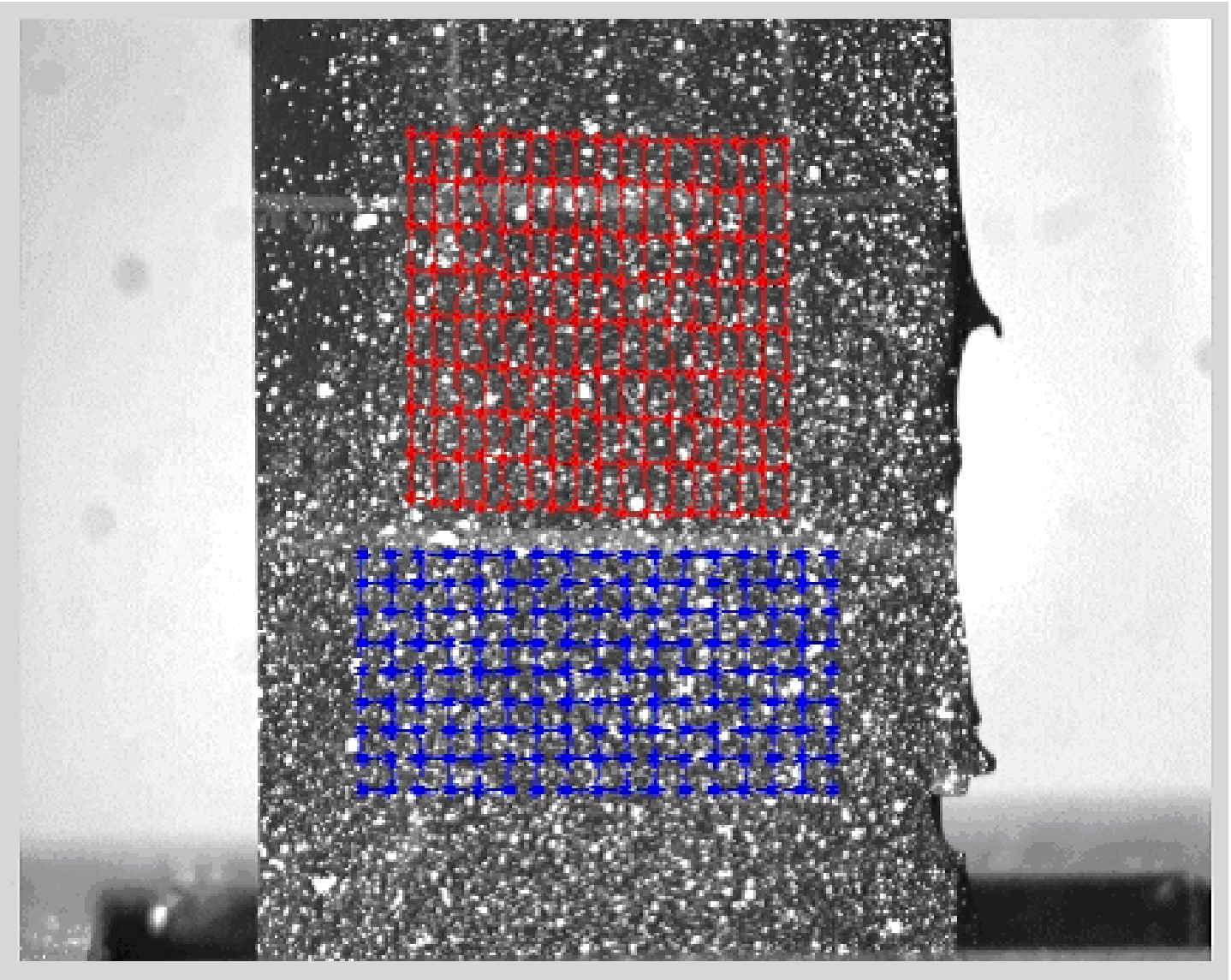

**Figure 3(a):** Large displacement measured with Correlli$^{GD}$ during in uni-axial tension test on NR 70 (in blue initial grid and, in red deformed grid).

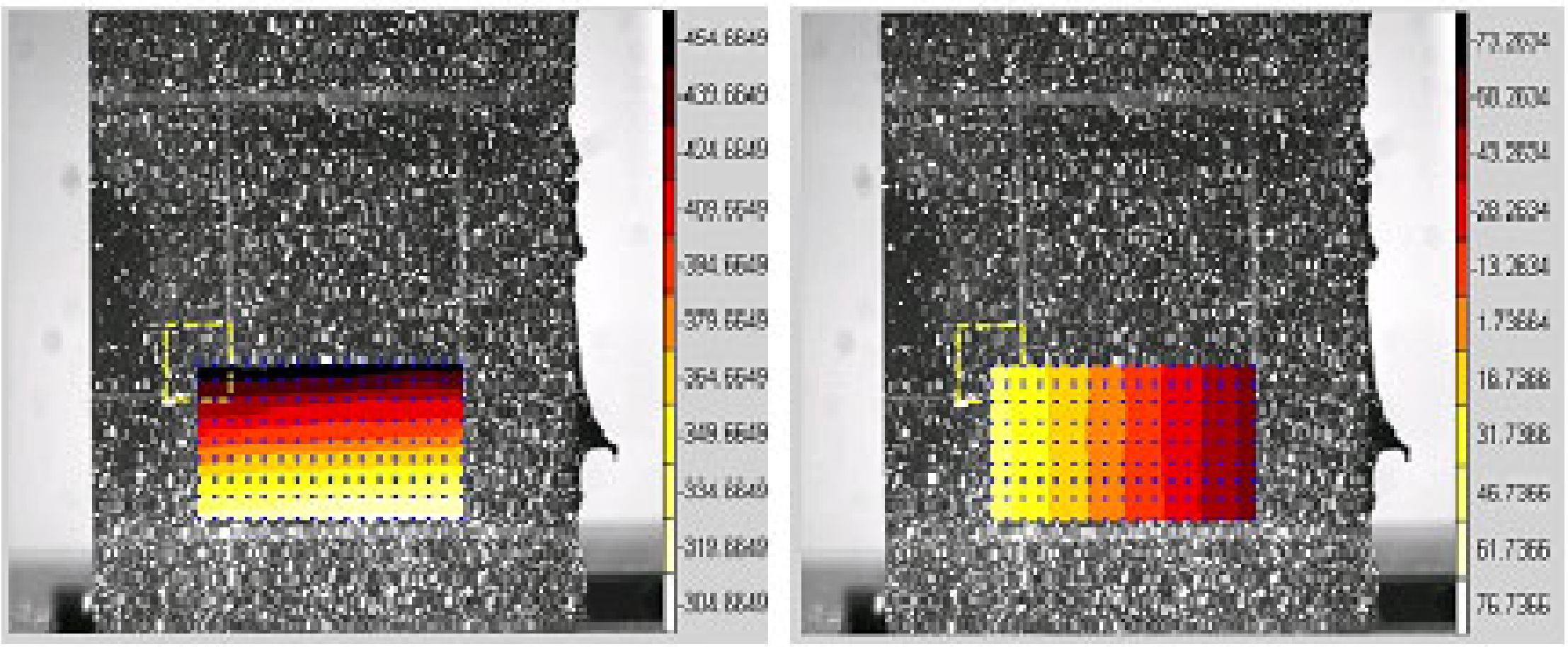

**Figure 3(b):** Displacements measured from image analysis on the NR 70, displacement components $U_{11}$ (left) and $U_{22}$ (right) are given in pixel. Parallel lines with longitudinal and transversal axis assure strain homogeneity of the studied zone

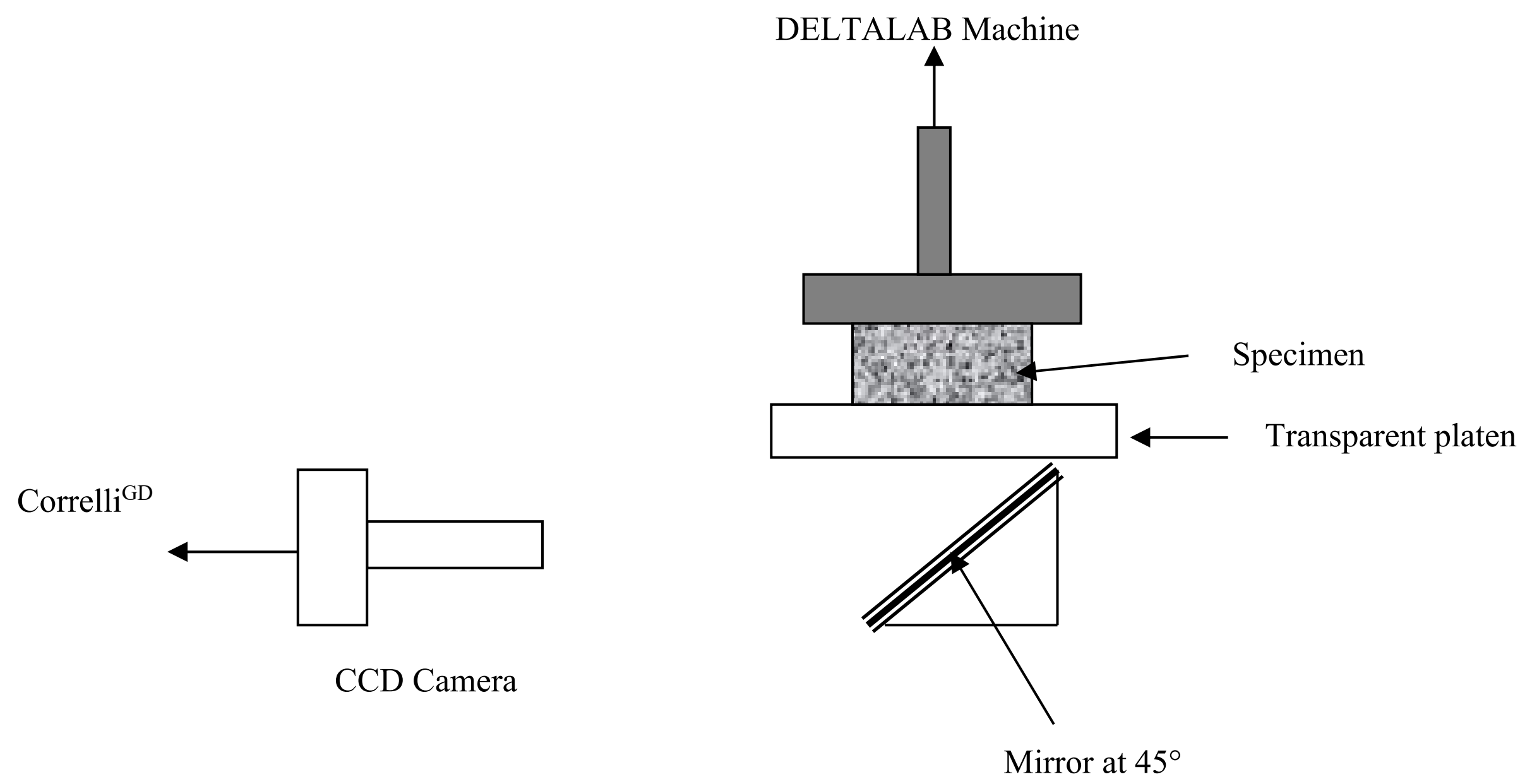

**Figure 4(a)**: Compression set up.

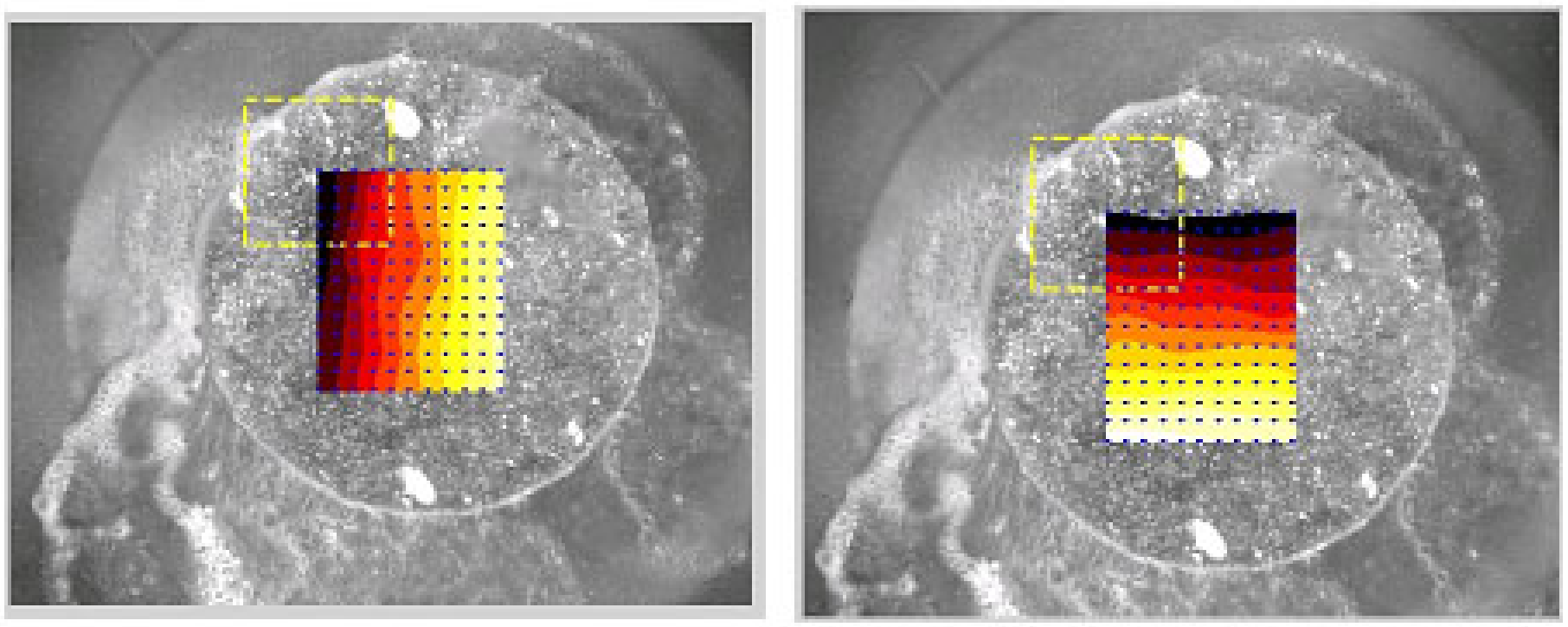

**Figure 4(b)**: Image analysis results and, radial distribution of the components of displacements components $U_{11}$ and $U_{22}$ .

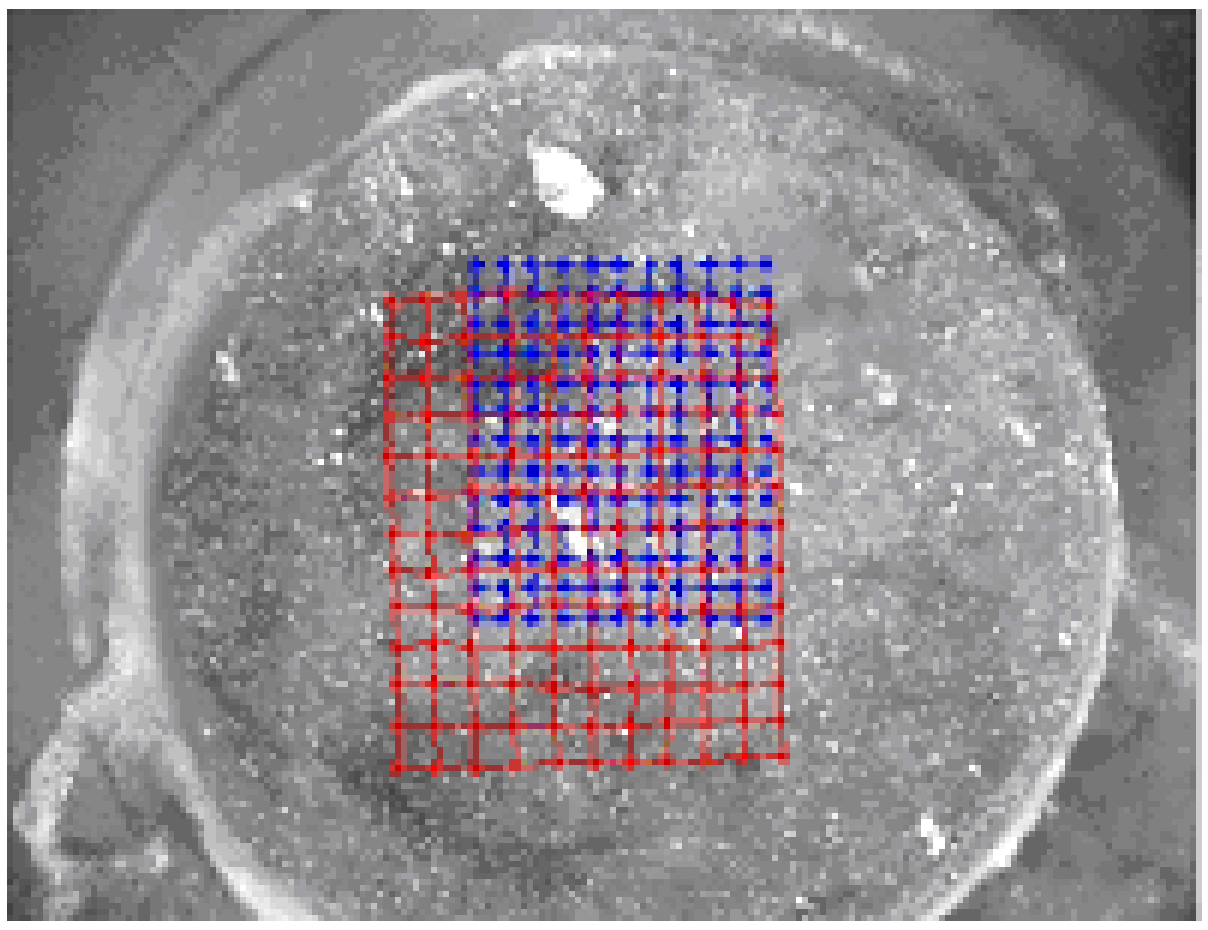

**Figure 4(c)**: Image analysis results and, radial distribution of the components of strain tensor ($\varepsilon_{11}$ and $\varepsilon_{22}$) in the simple compression test (in blue initial grid and, in red deformed grid).

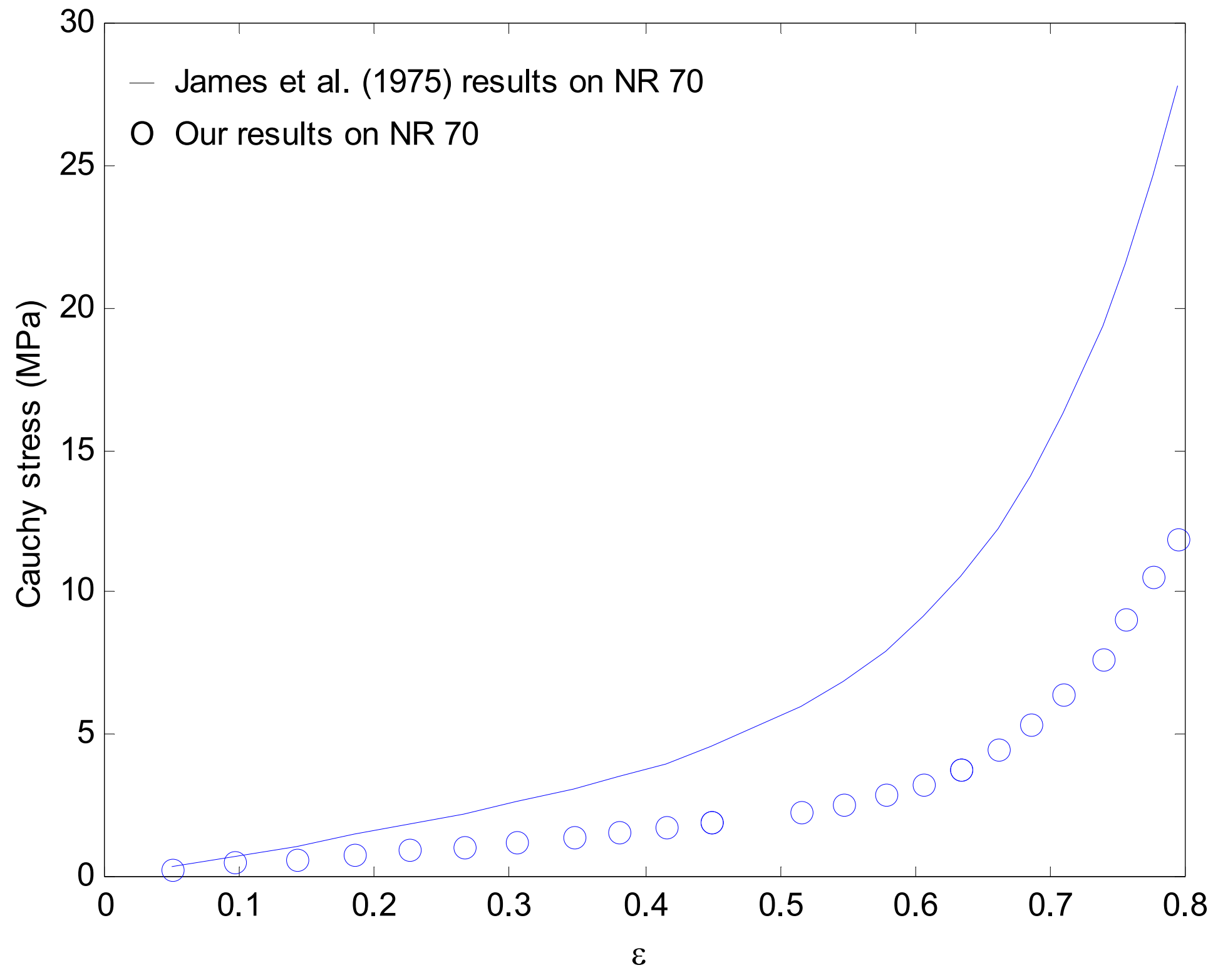

**Figure 5(a)**: Stress-elongation response of NR 70 in simple tension.

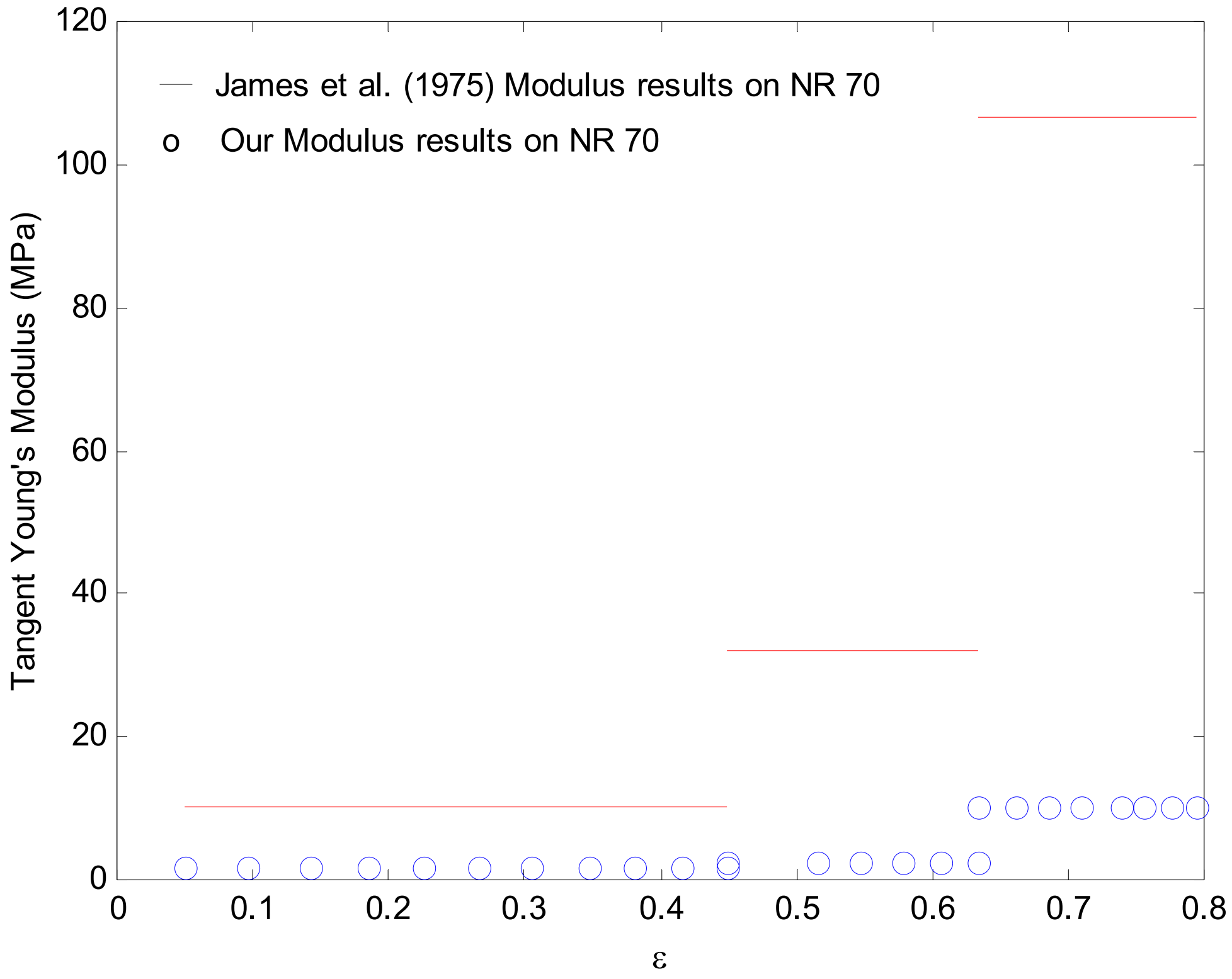

**Figure 5(b)**: Comparison of the tangent Young's Modulus of NR's 70 with strain obtained in simple tension test.

| Ingredient | James et al. (1975)<br>Material composite NR 70 | Our Material composite<br>NR 70 |
|---|---|---|
| Natural rubber (NR) | 100 | 400 |
| Sulfur | 2.5 | 10 |
| C.B.S. | 0.5 | 2 |
| Stearic acid | 2 | 8 |
| ZnO | 5 | 20 |
| Mineral oil | 5 | 20 |
| Monox ZA | 0.15 | |
| Nonox BLB | 1.7 | |
| HAF | 70 | |
| C.B.S. | | 2 |
| N375 (C.B.) | | 280 |
| Processing cure | 15-50 min at 135°C | 15 min at 190°C |

.

**Table I** : Composition of Carbon Black (C.B.)/Natural Rubber (NR) composites, the amount of C.B. 70% .

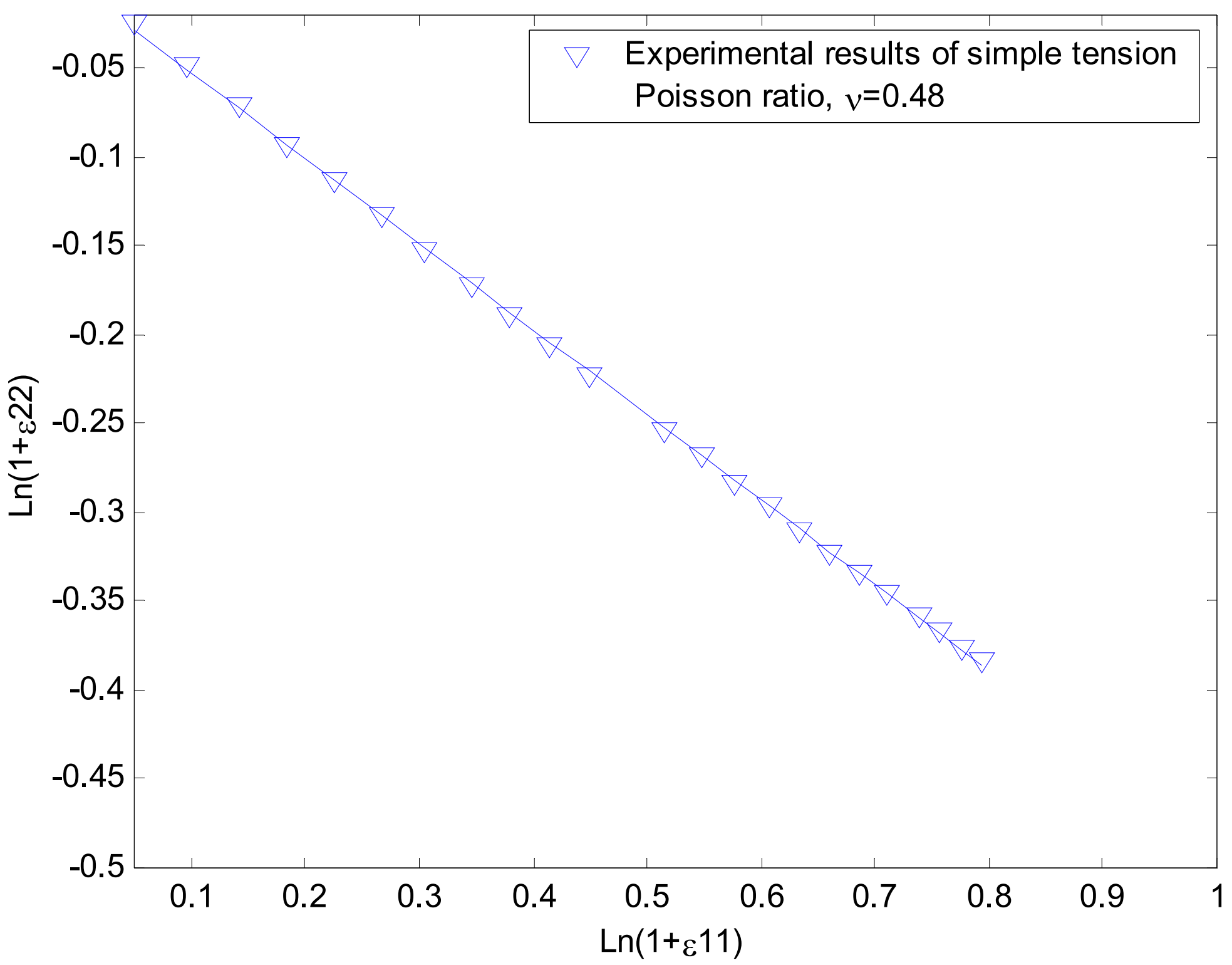

**Figure 6**: Logarithm of the simple lateral ratio versus Logarithm of the longitudinal extension.

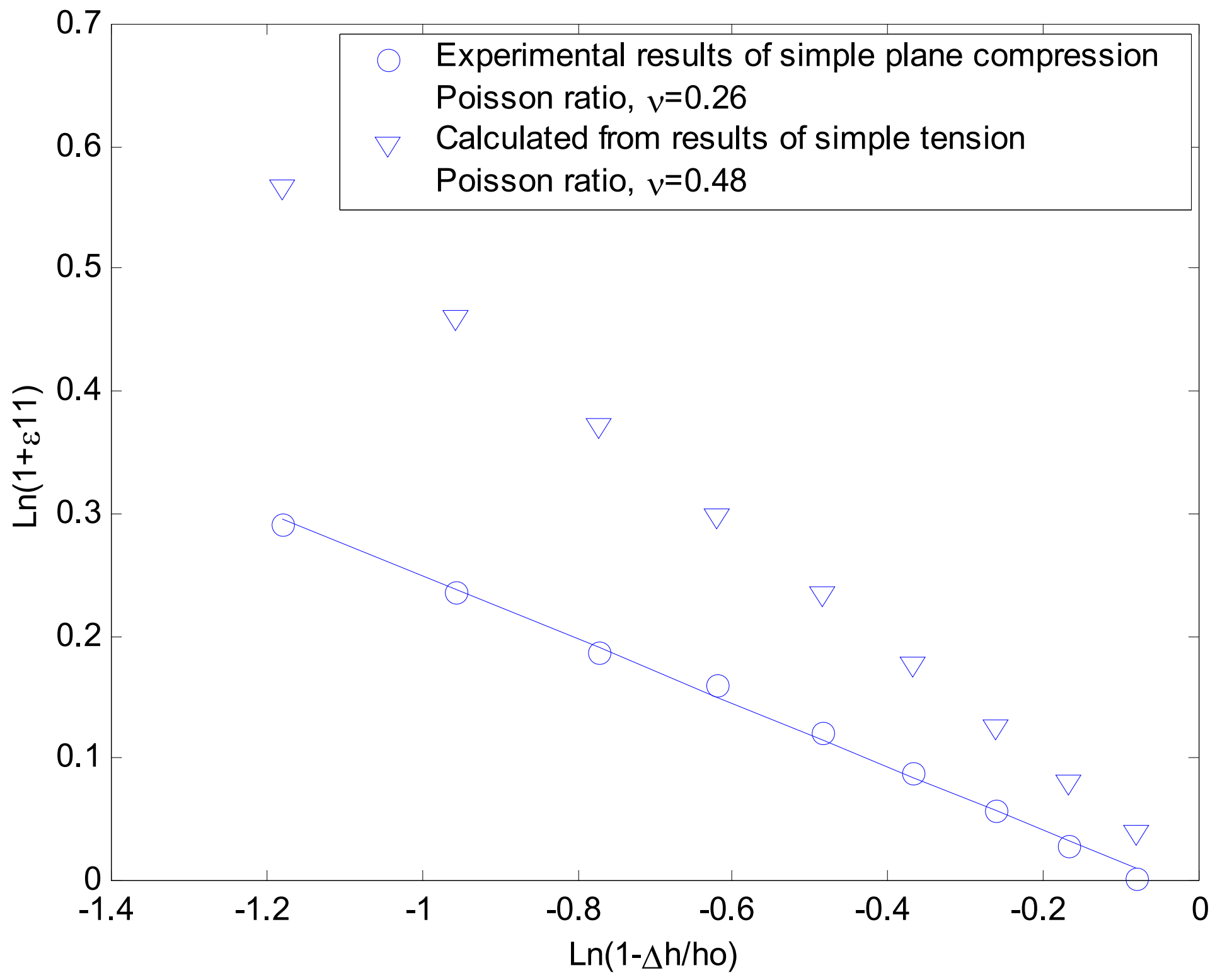

**Figure 7**: Comparison of the lateral elongation measured with the image analysis in simple plane compression and, the same results calculated from simple tension test, the values of the slope is different from the identified Poisson ratio.

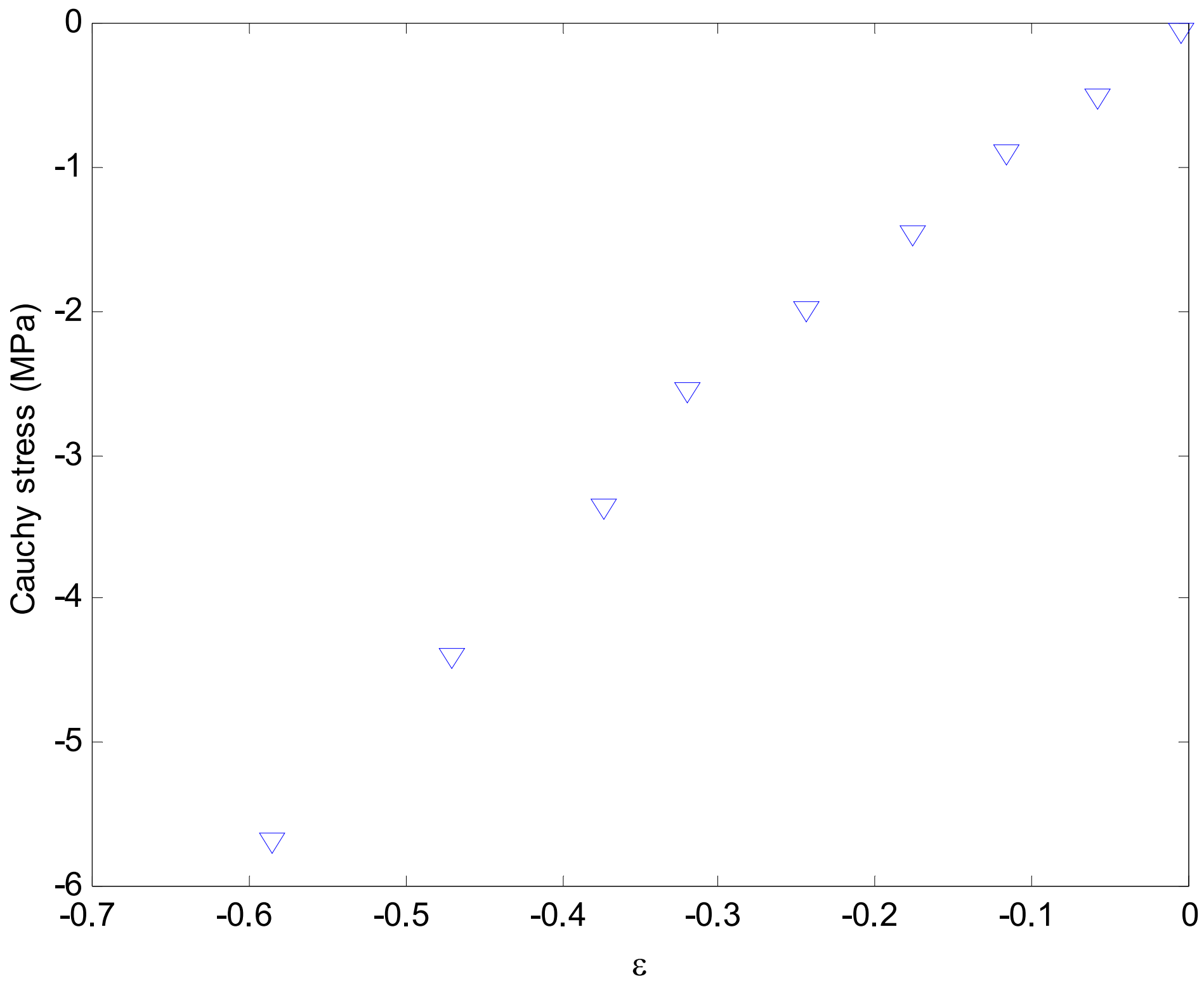

**Figure 8(a)**: Stress-strain of NR 70 in and simple plane compression, elongation calculated from the displacements of the traverse testing machine.

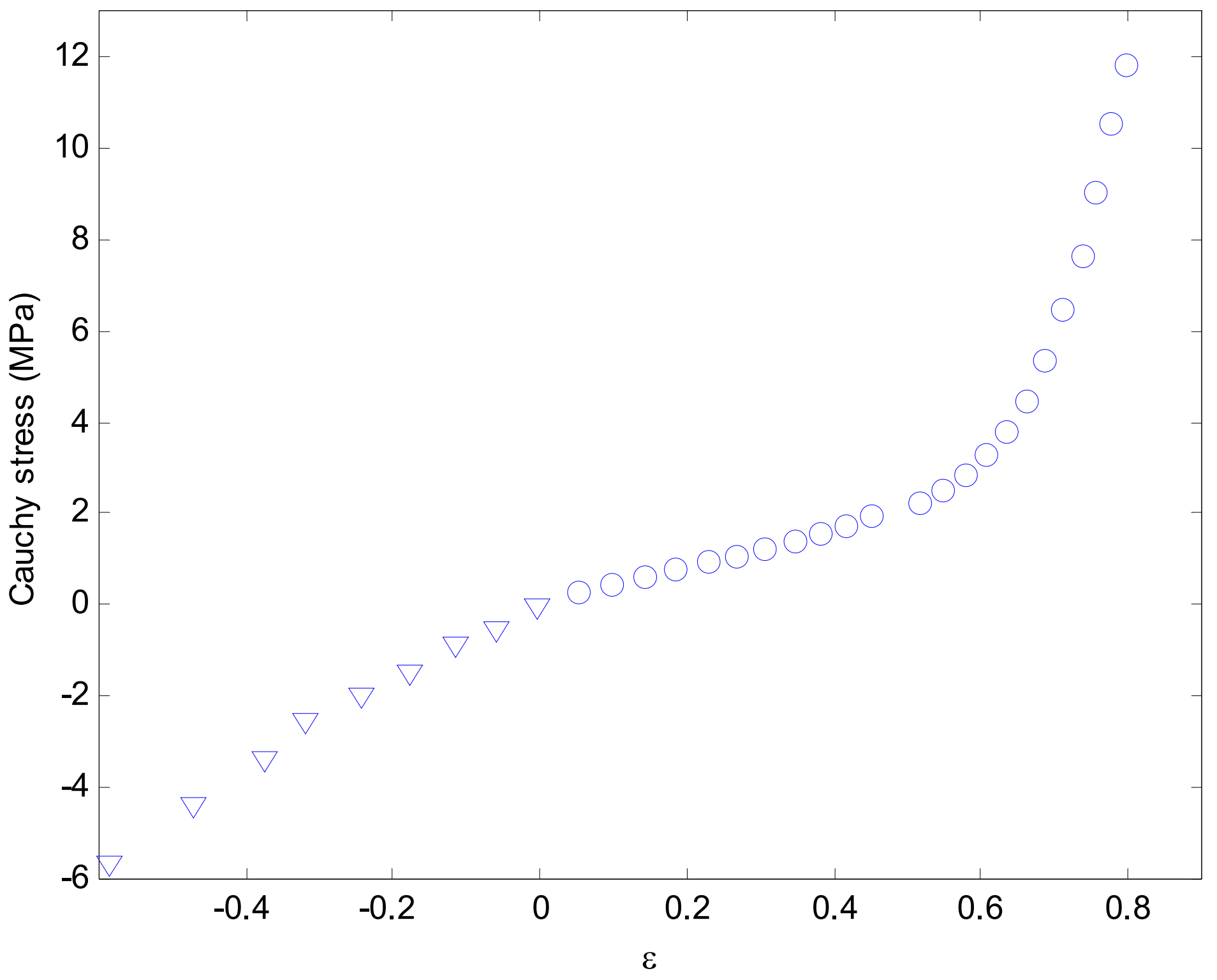

**Figure 8(b)**: Stress-strain response of NR 70 in simple plane compression and simple tension tests.

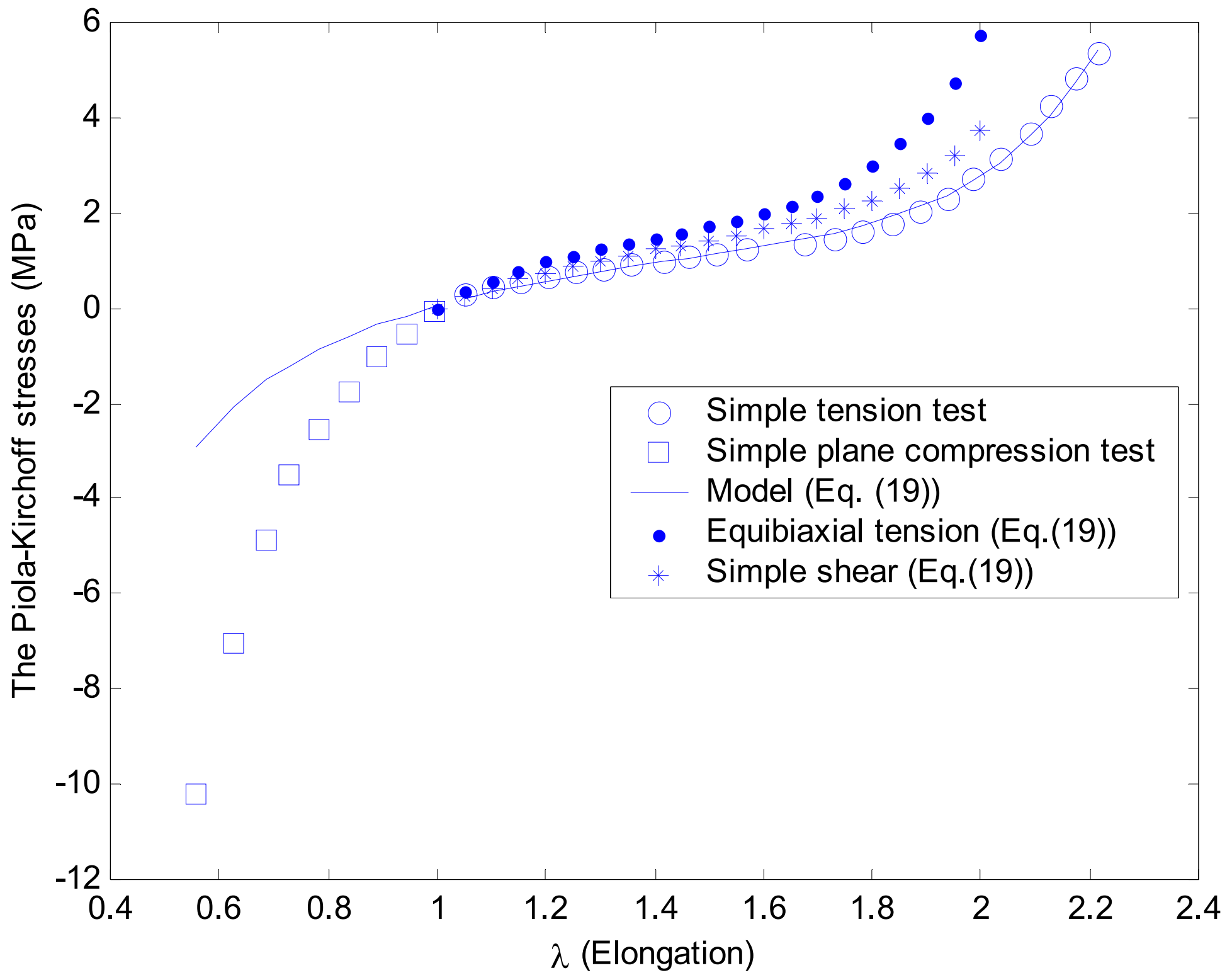

**Figure 9**: Stress-strain results measured in simple tension and in simple plane compression on NR 70, and the predictions of our Model.

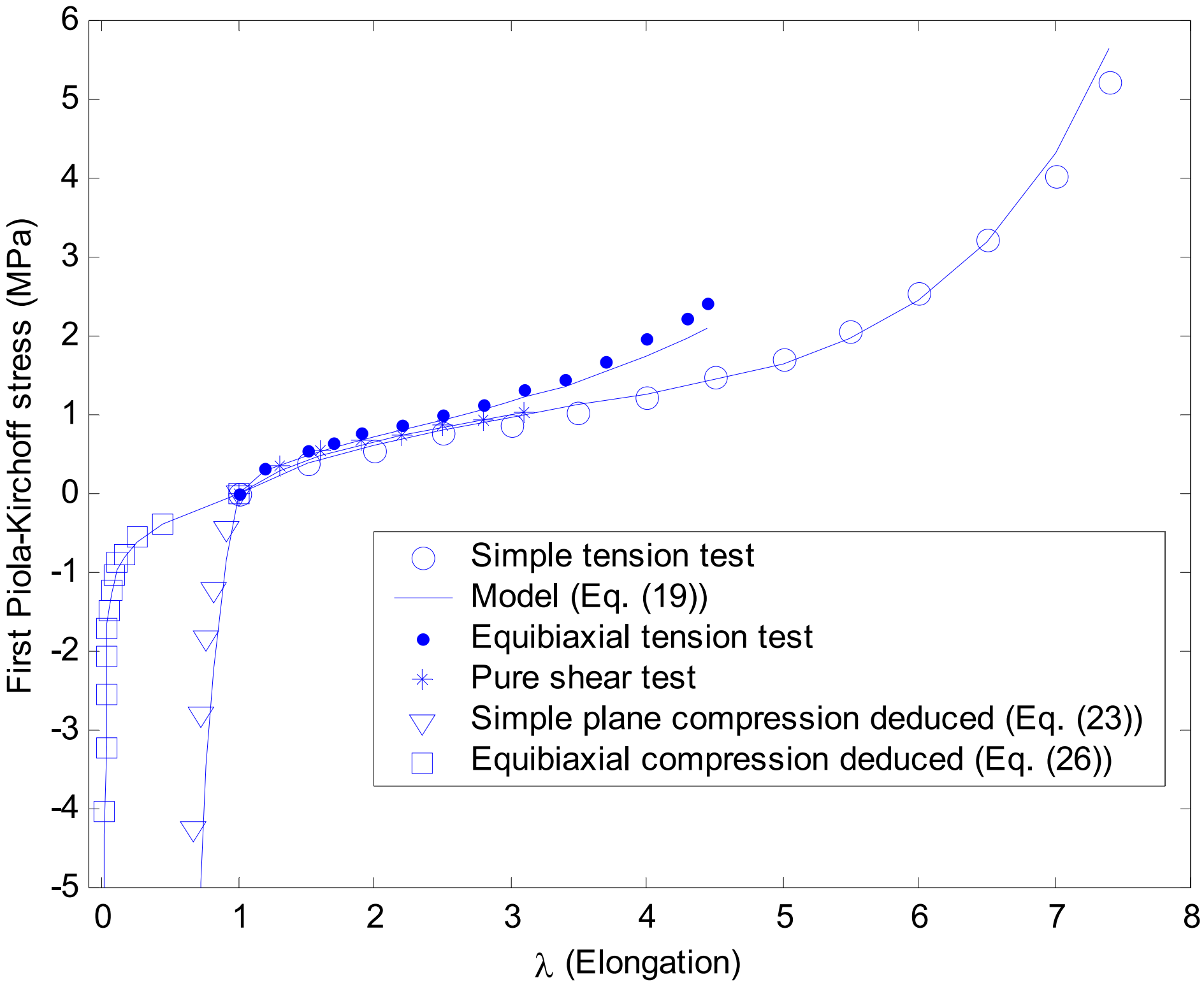

**Figure 10**: Comparison of the predicted results of our Model to the experimental results of Treolar (1944) in the case of 8% sulphur rubber.

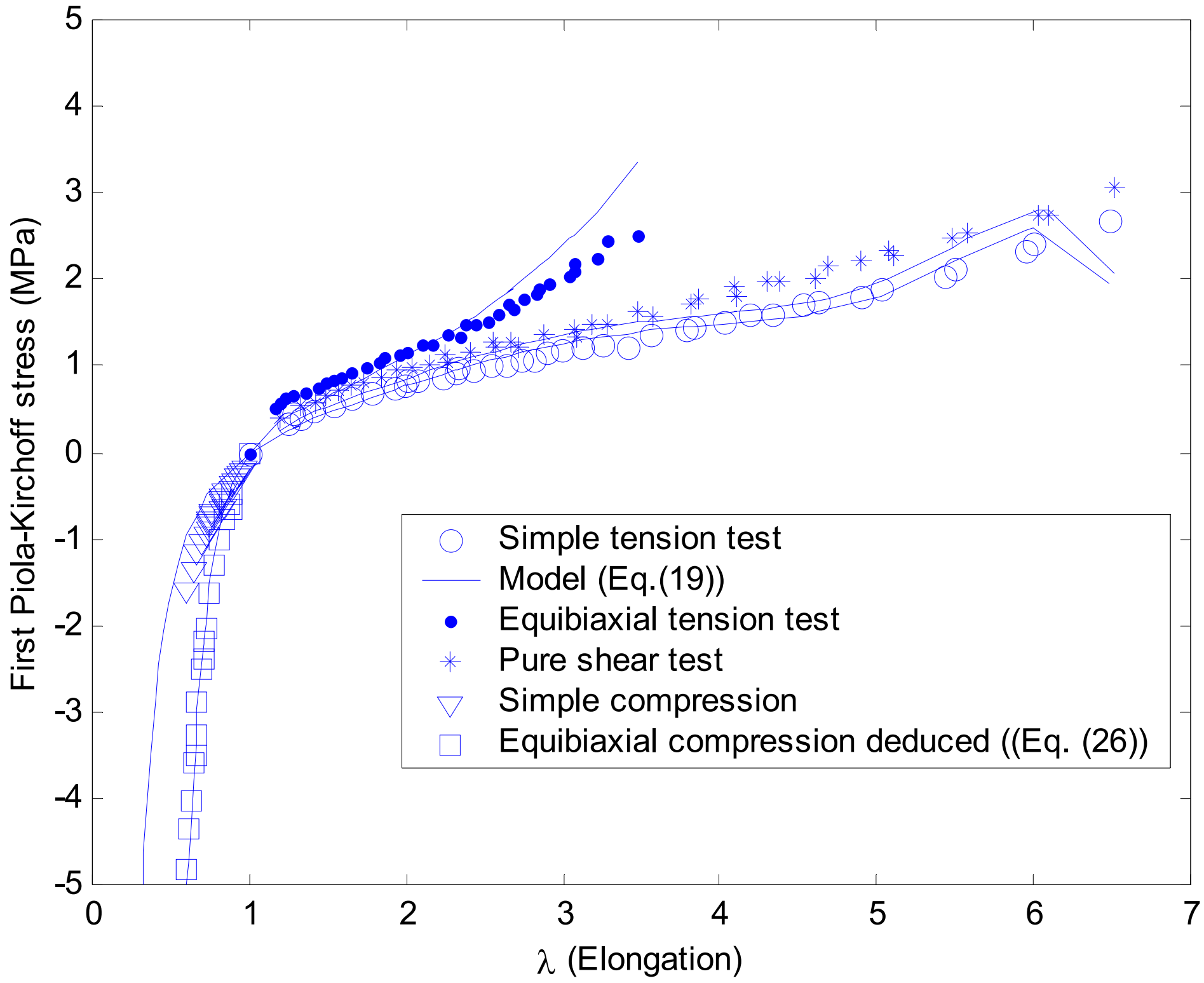

**Figure 11**: Comparison of the prediction of our Model to the experimental results of Heuillet *et al.* (1997) in the case of NR (Natural Rubber).

| | Data of Treolar (1944) for a 8% sulphur rubber | Results for NR 70 | Data of Heuillet et al. (1997) for a natural rubber . |
|---|---|---|---|
| Neo-Hookean Model<br>$W=C_1^1(I_1-3)$ | $C_1^1 = 0.18$ MPa | $C_1^1 = 3.462$ MPa | $C_1^1 = 0.223$ MPa |
| Model Eq. (19 b) | $C_1^1 = 0.18$ MPa<br>$C_2^1 = 0.0015$ MPa<br>$C_1^2 = -0.0023$ MPa<br>$C_2^2 = 0.044.10^{-5}$ MPa | $C_1^1 = 0.62582$ MPa<br>$C_1^2 = 0.000113$ MPa<br>$C_2^1 = -0.02070$ MPa<br>$C_2^2 = 0.000123$ MPa<br>$C_3^1 = 0.000002167$ MPa<br>$C_3^2 = 0.00004859$ MPa | $C_1^1 = 0.223$ MPa<br>$C_1^2 = 0.0087$ MPa<br>$C_2^1 = -0.0012$ MPa<br>$C_2^2 = 0.00001$ MPa<br>$C_3^1 = -0.00052$ MPa<br>$C_3^2 = -0.15.10^{-8}$ MPa |

**Table II**: Material parameters for rubbers (NR's) and NR 70.